\documentclass[aps,pre,reprint,superscriptaddress,amsmath,amssymb,showpacs,floatfix]{revtex4-1}
\usepackage{graphicx}
\usepackage{bm}
\usepackage{color}
\usepackage[applemac]{inputenc}
\usepackage[T1]{fontenc}

\begin{document}

\title{Thermodynamic uncertainty relation for underdamped dynamics driven by time-dependent protocols }

\author{Chulan Kwon}
\affiliation{Department of Physics, Myongji University, Yongin, Gyeonggi-Do,
17058,  Korea}
\email{ckwon@mju.ac.kr}
\author{Youngchae Kwon}
\affiliation{Department of Physics, Myongji University, Yongin, Gyeonggi-Do,
17058,  Korea}
\author{Hyun Keun Lee}
\affiliation{Department of Physics, Sungkyunkwan University, Suwon, Gyeonggi-do 16419, Korea}
\date{\today}

\begin{abstract}
The thermodynamic uncertainty relation (TUR) for underdamped dynamics has intriguing problems while its counterpart for overdamped dynamics has recently been derived. Even for the case of steady states, a proper way to match underdamped and overdamped TURs has not been found. We derive the TUR for underdamped systems subject to general time-dependent protocols, that covers steady states, by using the Cram\'{e}r-Rao inequality. We show the resultant TUR to give rise to the inequality of the product of the variance and entropy production.  We prove it to approach to the known overdamped result for large viscosity limit. We present three examples to confirm our rigorous result.
\end{abstract}
\maketitle

\emph{\textbf{Introduction}}. Non-equilibrium fluctuation in time-accumulated observables such as work, heat, and entropy production (EP) has been an important subject since the discovery of the fluctuation theorem (FT) for deterministic systems about two decades ago \cite{evans93,evans94,gallavotti}. The FT with various forms  has been proved to hold for stochastic systems theoretically \cite{jarzynski1,jarzynski2,crooks,kurchan,lebowitz,seifert,esposito} and experimentally \cite{wang,hummer,liphard,trepagnier,collin,garnier,hayashi,dongyun}. In recent years, another fundamental property on non-equilibrium fluctuation, the thermodynamic uncertainty relation (TUR), has been discovered \cite{barato1,gingrich1}. The TUR  provides a universal trade-off relation between the fluctuation of an arbitrary time-accumulated observable and its dissipation. For a non-equilibrium steady state, the TUR for an observable $\Phi$ accumulated over time $\tau$ states 
\begin{equation}
\frac{{\rm Var} ~\Phi}{\langle\Phi\rangle^2}\ge \frac{2 k_B}{\tau \sigma}~,
\label{TUR_ness}
\end{equation}
where ${\rm Var}~\Phi$ is the variance of $\Phi$, $\langle\Phi^2\rangle-\langle\Phi\rangle^2$, and $\sigma$ is the average EP rate. The proof is known for the systems with even parity in time reversal.

There have been extensive studies on the TUR for various systems in different dynamics. It has been studied for continuous-time Markov jumping processes for long-time \cite{gingrich1} and finite-time \cite{horowitz,gingrich2,pietzonka1}, also for discrete-time jumping process \cite{proesmans}, and for linear response systems \cite{macieszazak}. It has been studied for general overdamped Langevin systems \cite{dechant,hasegawa1} and heat engines \cite{pietzonka2}. For systems with time-dependent protocols, it has also been extensively studied for periodically driven systems in the absence of steady states \cite{barato2,barato3,koyuk1}. There have been alternative studies to consider a different type of inequality, the generalized TUR, in use of the FT \cite{hasegawa2,potts}. 

For general time-dependent protocols, the TUR has not been throughly confirmed. For overdamped Langevin systems, the rigorous derivation of the TUR has been found by Koyuk and Seifert \cite{koyuk2}, which states in terms of our notation
\begin{equation}
\frac{{\rm Var} ~\Phi}{\left[(\tau\partial_\tau-\omega\partial_\omega)\langle\Phi\rangle\right]^2}\ge \frac{2 k_B}{\langle \Delta S_{\rm tot}\rangle}~.
\label{TUR_OD}
\end{equation} 
where $\omega$ is the speed of the protocol change in time $t$, that appears as $\lambda(\omega t)$ for the protocol $\lambda$ and $\langle\Delta S_{\rm tot}\rangle$ is the average total EP produced in the system and bath. It holds for any systems prepared in arbitrary initial states and goes to the steady-state TUR in Eq.~\eqref{TUR_ness} in the absence of time-dependent protocol. There have been investigations for underdamped systems including momentum variables with odd-parity in time reversal \cite{fischer,chun,vu,lee}, but the recent result for steady states by Vu and Hasegawa \cite{vu} has left a controversy that it does not approach to the known overdamped TUR in Eq.~\eqref{TUR_OD} and even gives rise to a looser bound.  

In this study, we investigate the TUR for underdamped dynamics subject to general time-dependent protocols which goes to Eq.~\eqref{TUR_OD} for large viscosity. First, we will derive the TUR rigorously and show it to go to the proper overdamped limit. We will present a few examples to confirm our results.

\emph{\textbf{Derivation of the TUR}}. We consider a particle of mass $m$ in underdamped stochastic dynamics subject to a general time-dependent protocol. The probability distribution function (PDF), $\rho(\mathbf{x},\mathbf{p},t)$, for position $\mathbf{x}$ and momentum $\mathbf{p}$ satisfies the Kramers equation
\begin{equation}
\partial_t \rho(\mathbf{x},\mathbf{p},t)=-\partial_{\mathbf{x}}\cdot \mathbf{j}_{\mathbf{x}}-\partial_{\mathbf{p}}\cdot\mathbf{j}_{\mathbf{p}}
\label{Kramers}
\end{equation}
where $\partial_t$, $\partial_\mathbf{x}$, and $\partial_\mathbf{p}$ denote partial differentiations with respect to variables in subscript. $\mathbf{j}_{\mathbf{x}}=(\mathbf{p}/m )\rho$ and $\mathbf{j}_{\mathbf{p}}=[\mathbf{H}(\mathbf{x},\mathbf{p},t)-D\partial_{\mathbf{p}}]\rho$ for the drift term $\mathbf{H}=-(\gamma/m)\mathbf{p}+\mathbf{f}(\mathbf{x},\omega t)$ are the probability current densities in position and momentum spaces, respectively.
$\gamma$ and $D$ are the viscosity and diffusion coefficients, respectively, satisfying the Einstein relation, $D=\gamma\beta^{-1}$ for the inverse temperature $\beta$ of the medium (bath) maintained in equilibrium. For simplicity of notation, we consider a single colloidal particle in isotropic medium, which can be extended to a many-particle system in non-isotropic medium with viscosity and diffusion coefficients in matrix forms. The time-dependent protocol is included in the force $\mathbf{f}$ and $\omega$ is introduced as a parameter to control the speed of protocol change in time \cite{koyuk2}; for example, it is the frequency of an oscillating driving force. The force can be given by $\mathbf{f}=-\partial_{\mathbf{x}}V(\mathbf{x}, \omega t)$ for time-the dependent protocol in a potential $V$ or $\mathbf{f}=-\partial_{\mathbf{x}}V(\mathbf{x})+\mathbf{f}_{\rm nc}(\mathbf{x},\omega t)$ for the time-dependent protocol in a nonconservative external force. In the absence of time-dependent protocol, the system reaches a thermal equilibrium with the Boltzmann distribution, $\rho\propto e^{-\beta E}$ for the energy $E=p^2/(2m)+V(\mathbf{x})$.

We consider a time-accumulated production $\Phi[\mathbf{q}(t)]$ which is a functional of path $\mathbf{q}(t)=(\mathbf{x}(t),\mathbf{p}(t))$ for $0\le t\le \tau$ in the phase space. We write the production rate $\dot{\Phi}$ per time written in the form of either $\partial_t\chi(\mathbf{x},\omega t)$ (type I) or $\mathbf{g}(\mathbf{x},\omega t)\cdot\dot{\mathbf{x}}$ (type II). For example, the work production rate is given by $\partial_t V(\mathbf{x},\omega t)$ or $\mathbf{f}_{\rm nc}(\mathbf{x},\omega t)\cdot\dot{\mathbf{x}}$. The average total production $\langle \Phi(\mathbf{q}(t)\rangle$ over many trajectories can be computed in principle by the path integral with the path probability 
\begin{equation}
P[\mathbf{q}(t)]\propto \rho_{\rm in}(\mathbf{q}_0)\prod_t\delta (\dot{\mathbf{x}}-\mathbf{p}/m) e^{-\int_0^\tau dt {\cal L}}
\label{path_prob}
\end{equation}
where the Lagrangian density is given by ${\cal L}=(1/4)D^{-1}[\dot{\mathbf{p}}-\mathbf{H}]^2+\epsilon\partial_{\mathbf{p}}\cdot\mathbf{H}$ by using a discretization parametrized by $0\le \epsilon\le 1$ where $\dot{\mathbf{p}}\simeq [\mathbf{p}(t)-\mathbf{p}(t-dt)]/dt$ and $\mathbf{H}\simeq \epsilon \mathbf{H}(t)+(1-\epsilon)\mathbf{H}(t-dt)$ \cite{kwon_jkps} though the path integral is independent of discretizations; typically, $\epsilon=0$ (prepoint, Ito), $1/2$ (midpoint, Stratonovich), $1$ (postpoint, anti-Ito).

We follow the recent approach via the Cram\'{e}r-Rao inequality to derive the TUR~\cite{hasegawa1,vu}. It is crucial to choose the adjoint dynamics which differs from the original one by the perturbation parameter $\theta$. Out of many choices for the adjoint dynamics, we propose one to give rise to the TUR which relates the variance of $\Phi$ and the total entropy production (EP), and goes to that for overdamped dynamics for large $\gamma$ recently derived by Koyuk and Seifert \cite{koyuk2}. The adjoint dynamics is governed by the chosen drift term
\begin{equation}
\mathbf{H}_\theta =-\frac{\gamma}{m(1+\theta)}\mathbf{p}+\alpha\mathbf{f}(\mathbf{x},\omega t)+\left(1-\frac{1}{1+\theta}\right)D\partial_{\mathbf{p}}\ln \rho_\theta
\label{H_theta}
\end{equation}
where $\rho_\theta$ is the PDF of the adjoint dynamics. $\alpha$ is introduced for a later use, which is set to be unity for the time being. Then, Eqs.~\eqref{Kramers} and \eqref{path_prob} are modified by $\mathbf{H}_\theta$. The Cram\'{e}r-Rao inequality is written as
\begin{equation}
\frac{{\rm Var}_\theta \Phi }{[\partial_\theta \langle \Phi\rangle_\theta]^2}\ge \frac{1}{{\cal I}(\theta)}
\label{Cramer-Rao}
\end{equation}
where ${\rm Var}_\theta \Phi=\langle (\Phi-\langle\Phi\rangle)^2\rangle_\theta$ is the variance of the production and ${\cal I}(\theta)=-\langle\partial_\theta^2 \ln P_\theta[\mathbf{q}]\rangle$ is the Fischer information. It goes to a TUR in the limit $\theta\to 0$. 

The Fischer information ${\cal I}(0)$ is found from Eq.~\eqref{path_prob}. Using the midpoint (Stratonovich) representation with $\epsilon=1/2$,
\begin{eqnarray}
{\cal I}(0)&=&\frac{D^{-1}}{2}\int_0^\tau dt \left\langle(\partial_\theta \mathbf{H}_\theta)^2-(\dot{\mathbf{p}}-\mathbf{H}_\theta)\cdot \partial_\theta^2\mathbf{H}_\theta\right.
\nonumber\\
&&\left.+D\partial_\theta^2\partial_\mathbf{p}\cdot \mathbf{H}_\theta\right\rangle\Big|_{\theta=0}
\end{eqnarray}
where the second and third terms are shown to be cancelled by using the property of midpoint discretization that $\langle \partial_\theta^2\mathbf{H}_\theta\cdot(\dot{\mathbf{p}}-\mathbf{H})\rangle=\langle \partial_\theta^2\mathbf{H}\cdot(\mathbf{j}_\mathbf{p}/\rho-\mathbf{H})\rangle$ and the integration by parts
$\langle\partial_\theta^2\mathbf{H}\cdot D\partial_\mathbf{p}\rho/\rho\rangle=-\langle D\partial_\mathbf{p}\cdot\partial_\theta^2\mathbf{H}_\theta\rangle$. We find $[\partial_\theta\mathbf{H}_\theta]_{\theta=0}= (\gamma/m)\mathbf{p}+D\partial_{\mathbf{p}}\ln\rho=-\mathbf{j}^{\rm irr}_\mathbf{p}/\rho$ where $\mathbf{j}^{\rm irr}_\mathbf{p}$ is the irreversible current density in the momentum space, known as the irreversibility measure for non-equilibrium  \cite{risken,spinney,yeo}. Hence, we have 
\begin{equation}
{\cal I}(0)=\frac{D^{-1}}{2}\int_0^\tau dt\left \langle \left(\mathbf{j}^{\rm irr}/\rho\right)^2\right\rangle=\frac{1}{2 k_B}\langle \Delta S_{\rm tot}\rangle
\end{equation}
where  $\langle\Delta S_{\rm tot}\rangle$ is the average total EP in the system and bath. Note that ${\cal I}(0)$ from the different choice of $\mathbf{H}_\theta$ by Vu and Hasegawa has extra terms besides  the average EP, which remains even for the overdamped limit. 

The average total production is given by $\langle \Phi\rangle_\theta=\int_0^\tau dt \int d\mathbf{x}\int d\mathbf{p} \Gamma(\mathbf{x},\mathbf{p}, \omega t)\rho_\theta(\mathbf{x},\mathbf{p},t)$ where the rate function $\Gamma$ associated with $\dot\Phi$ is independent of $\theta$ and has either form  
\begin{equation}
\Gamma(\mathbf{x},\mathbf{p}, \omega t)=\left\{\begin{array}{ll} \partial_t\chi(\mathbf{x},\omega t)& (\textrm{type I}) \\ \mathbf{g}(\mathbf{x},\omega t)\cdot\mathbf{p}/m & (\textrm{type II})
\end{array}\right.
\end{equation}
With $\mathbf{H}_\theta$ in Eq.~\eqref{H_theta}, $\rho_\theta$ satisfies the same form of the Kramers equation with renamed viscosity coefficient as $\bar{\gamma}=\gamma/(1+\theta)$. Then, we find 
\begin{equation}
 \langle \Phi\rangle_\theta=\int_0^{\tau}\!\!\! dt~\phi(t;\bar{\gamma})~,
 \label{Phi_theta}
 \end{equation}
where $\phi=\int d\mathbf{x}\int d\mathbf{p} \Gamma(\mathbf{x}, \mathbf{p},\omega t)\rho(\mathbf{x},\mathbf{p},t; \bar{\gamma})$. Therefore, we get
\begin{equation}
\partial_\theta\langle\Phi\rangle_\theta\Big|_{\theta=0}=-\gamma\partial_\gamma\langle\Phi\rangle~,
\label{D_Phi}
\end{equation}
Equation~\eqref{Cramer-Rao} gives the TUR 
\begin{equation}
\frac{{\rm Var} \Phi}{\left[-\gamma\partial_\gamma \langle\Phi\rangle\right]^2}\ge \frac{2 k_B}{\langle\Delta S_{\rm tot}\rangle}~.
\label{TUR_UND}
\end{equation}
It is the rigorous result, while it leaves a practical problem that $\gamma$ might not be a proper control parameter in real experiments.

In order to express Eq.~\eqref{D_Phi} in terms of experimentally controllable parameters, we consider $H_\theta$ with $\alpha\neq 1$ in Eq.~\eqref{H_theta}. We introduce the change in variables and parameters such that $\bar{t}=t/(1+\theta)$, $\bar{\omega}=(1+\theta)\omega$, $\bar{\mathbf{p}}=(1+\theta)\mathbf{p}$, $\bar{\alpha}=(1+\theta)^2\alpha$, and $\bar{\beta}=(1+\theta)^{-2}\beta$. In this change, $\omega t$ remains having the same form as $\bar{\omega}\bar{t}$. The resultant Kramers equation has an invariant form with these renamed variables and parameters. 
The initial PDF of the original system is written as $\rho_{\rm in}(\mathbf{x}_0,\mathbf{p}_0; \{\lambda_i\})$ with its own parameters, $\{\lambda_i\}$. Then, the corresponding initial PDF of the adjoint system is given by $\rho_\theta(\mathbf{x},\mathbf{p},0)
=(1+\theta)^d \rho_{\rm in}(\mathbf{x},\bar{\mathbf{p}}; \{\bar{\lambda}_i\})$ where $\bar{\lambda}_i$s are renamed parameters depending on the detailed form of the initial PDF and $d$ is the dimensionality. As a result, we have
\begin{eqnarray}
\rho_\theta(\mathbf{x},\mathbf{p},t)&=&(1+\theta)^d\int d\mathbf{x}_0 d\bar{\mathbf{p}}_0{\cal T}(\mathbf{x},\bar{\mathbf{p}},\bar{t}|\mathbf{x}_0,\bar{\mathbf{p}}_0) \nonumber\\
&& \times\rho_{\rm in}\left(\mathbf{x}_0,\bar{\mathbf{p}}_0; \{\bar{\lambda}_i\}\right)~,
\end{eqnarray}
where ${\cal T}$ is the transition probability which is the solution of the renamed Kramers equation. It depends on renamed parameters, $\bar{\omega}$, $\bar{\alpha}$, and $\bar{\beta}$, included in ${\cal T}$ and also on initial parameters $\{\bar{\lambda}_i\}$. It is subtle to take into account the dependence of the initial condition on $\theta$. 
There are special but practical initial PDFs realized in experiments. One can usually prepare a system in equilibrium or local equilibrium with an initial PDF 
\begin{equation}
\rho_{\rm in}\propto e^{\beta\alpha V(\mathbf{x}-\mathbf{a},0)+\beta (\mathbf{p}-m\mathbf{v})^2/(2m)}=e^{\bar{\beta}\bar{\alpha}V+\bar{\beta}(\bar{\mathbf{p}}-m\bar{\mathbf{v}})^2/(2m)}~.
\label{initial}
\end{equation}
Note that $\bar{\mathbf{v}}=(1+\theta)\mathbf{v}$ for the initial average velocity $\mathbf{v}$ is the only additional renamed parameter to be considered.

For those conditions, we find 
$\rho_\theta=\rho(\mathbf{x},\bar{\mathbf{p}},\bar{t}; \bar{\alpha}, \bar{\beta},\bar{\omega},\bar{\mathbf{v}})(1+\theta)^d$. The rate function reads $\Gamma(\mathbf{x},\mathbf{p},\omega t)=\Gamma(\mathbf{x},\bar{\mathbf{p}},\bar{\omega}\bar{t})/(1+\theta)$ for which we use the scaling such that  $\partial_t\chi=\partial_{\bar{t}}\chi/(1+\theta)$ (type I) and $\mathbf{g}\cdot\mathbf{p}/m=(\mathbf{g}\cdot
\bar{\mathbf{p}}/m)/(1+\theta)$ (type II). Then, we find 
\begin{equation}
 \langle \Phi\rangle_\theta=\int_0^{\tau/(1+\theta)}\!\!\! d\bar{t }~\bar{\phi}(\bar{t};\bar{\alpha}, \bar{\beta},\bar{\omega},\bar{\mathbf{v}})
 \label{Phi_theta}
 \end{equation}
where $\bar{\phi}=\int d\mathbf{x}\int d\bar{\mathbf{p}}\Gamma(\mathbf{x},\bar{\mathbf{p}}, \bar{\omega}\bar{t})\rho(\mathbf{x},\bar{\mathbf{p}},\bar{t}; \bar{\alpha}, \bar{\beta},\bar{\omega},\bar{\mathbf{v}})$. Then, we find 
\begin{equation}
-\gamma\partial_\gamma\langle\Phi\rangle=\left(-\tau\partial_\tau+\omega\partial_\omega+2\alpha\partial_\alpha-2\beta\partial_\beta+\mathbf{v}\cdot\partial_\mathbf{v}\right)\langle\Phi\rangle~,
\label{D_Phi_new}
\end{equation}
where $-\beta\partial_\beta$ can be replaced by $T\partial_T$ for the temperature $T$ of the bath. The left-hand-side is useful for theoretical studies, while the right-hand-side is measurable in experiments. Note that this expression is valid for the practical initial condition given in Eq.~\eqref{initial}. 

%Changing the variable as $\bar{t}=t/\bar{\tau}_p$, we get
%$
%\langle\Phi\rangle_\theta =\int_0^{\tau/\bar{\tau}_p} d\bar{t}~ \hat{\phi}(\bar{t},\bar{\tau}_p,\bar{\omega})
%$
%where $\hat{\phi}(\bar{t},\bar{\tau}_p,\bar\omega)=\bar{\tau}_p \phi(\bar{t}\bar{\tau}_p,\bar{\tau}_p,\bar{\omega}/\bar{\tau}_p)$ with $\bar{\omega}=\omega\bar{\tau}_p$. Then, we can rewrite
%\begin{equation}
%\tau_p\partial_{\tau_p}\langle\Phi\rangle=(-\tau\partial_\tau+\omega\partial_\omega)\langle\Phi\rangle+K(\tau,\tau_p,\omega)
%\end{equation}
%where 
%$K(\tau,\tau_p,\omega)=\int_0^{\tau/\tau_p} d\bar{t}~\tau_p\partial_{\tau_p}\hat{\phi}(\bar{t},\tau_p,\bar{\omega}=\omega\tau_p).$

\emph{\textbf{Overdamped limit}}. We can show that the found TUR for large $\gamma$ goes to that recently derived by Koyuk and Seifert for the overdamped dynamics. The rigorous proof is possible by using the PDF form for small  $\gamma^{-1}$ obtained from an earlier work by one of us \cite{ao}, given as
\begin{equation}
\rho(\mathbf{x},\mathbf{p},t)\propto \exp\left[ -(1+g)\frac{\beta}{2m}\left(\mathbf{p}-m\mathbf{u}\right)^2-{\cal V}(\mathbf{x},t)\right]~.
\label{OD_PDF}
\end{equation}
Here, ${\cal V}$ is the leading-order potential landscape, $-\ln \rho(\mathbf{x},t)$, for the PDF in the position space given by  $\int d\mathbf{p} \rho(\mathbf{x},\mathbf{p},t)$. $g=(m/\gamma)\partial_\mathbf{x}\cdot\mathbf{u}$ gives the lowest-order correction to the overdamped limit, which we will neglect in the following derivation. $\mathbf{u}=\gamma^{-1}[\alpha\mathbf{f}+\beta^{-1}\partial_\mathbf{x}{\cal V}]$ is the velocity field of the probability current in the position space, $\mathbf{j}_\mathbf{x}=\mathbf{u}\rho(\mathbf{x},t)$. This equation manifests our expectation that fast-varying velocity variable maintains a local equilibrium distribution around an instantaneous average velocity $\mathbf{u}$ along the trajectory of slowly varying position variable. Averaging the Kramers equation~\eqref{Kramers} over momentum, we get the expected Fokker-Planck equation 
\begin{equation}
\partial_t \rho(\mathbf{x},t)=-\gamma^{-1}\partial_\mathbf{x}\cdot\left[\alpha\mathbf{f}-\beta^{-1}\partial_\mathbf{x}\right]\rho(\mathbf{x},t)~.
\end{equation}
Note that $\mathbf{j}_\mathbf{p}=-(\beta^{-1}\partial_\mathbf{x}{\cal V})\rho(\mathbf{x},\mathbf{p},t)$ using Eq.~\eqref{OD_PDF}, so $\int d\mathbf{p} ~\partial_\mathbf{p}\cdot\mathbf{j}_\mathbf{p}=0$. Changing variable as $s=t \alpha/\gamma$, $ \rho(\mathbf{x},t)=\rho(\mathbf{x},s; (\alpha\beta)^{-1},\omega\gamma\alpha^{-1})$. Note that and $\langle\Gamma(\mathbf{x},\mathbf{p}, \omega t)\rangle_{\mathbf{p}}=\alpha\gamma^{-1} \partial_s\chi$ (type I) or $\mathbf{g}\cdot \langle\mathbf{p}/m\rangle_{\mathbf{p}}=\alpha\gamma^{-1}\mathbf{g}\cdot[\mathbf{f}+(\alpha\beta)^{-1}\partial_\mathbf{x}{\cal V}]$ (type II) where $\langle \cdot\rangle_{\mathbf{p}}$ denotes the average over $\mathbf{p}$ withe the PDF in Eq.~\eqref{OD_PDF}. Wring $\langle\Gamma(\mathbf{x},\mathbf{p}, \omega t)\rangle_{\mathbf{p}}=\alpha\gamma^{-1}\bar{\Gamma}(\mathbf{x},(\omega\alpha\gamma^{-1})s)$, the equation \eqref{Phi_theta} for $\theta=0$ is given by 
\begin{equation}
 \langle \Phi\rangle=\int_0^{\tau\alpha\gamma^{-1}}\!\!\! ds~\phi^{\rm OD}(s;(\alpha\beta)^{-1},\omega\alpha\gamma^{-1})~,
\end{equation}
where $\phi^{\rm OD}=\int d\mathbf{x}\bar{\Gamma}(\mathbf{x},(\omega\alpha\gamma^{-1})s)\rho(\mathbf{x},s; (\alpha\beta)^{-1},\omega\alpha\gamma^{-1})$. Then, it is straightforward to find that Eq.~\eqref{D_Phi} up to the leading order in $\gamma^{-1}$ leads to 
\begin{equation}
-\gamma\partial_\gamma\langle\Phi\rangle=\left(\tau\partial_\tau-\omega\partial_\omega\right)\langle\Phi\rangle~.
\end{equation}
Note that the various terms in the right-hand-side of Eq.~\eqref{D_Phi_new} with no $\mathbf{v}$-derivative can be shown to collapse to the above simple form. It is indeed equal to $\partial_\theta\langle\Phi\rangle_\theta\Big|_{\theta=0}$ in the overdamped TUR for time-dependent protocols derived by Koyuk and Seifert, as seen in Eq.~\eqref{TUR_OD}.

It is pedagogically interesting to see $\langle\Delta S_{\rm tot}\rangle$ approach to that for the overdamped limit.
The irreversible current is found as $\mathbf{j}^{\rm irr}_\mathbf{p}=(-\gamma\mathbf{p}/m-D\partial_\mathbf{p})\rho=-\gamma\mathbf{u}\rho(\mathbf{x},\mathbf{p},t)$ in the leading order in $\gamma^{-1}$ by using Eq.~\eqref{OD_PDF}. Then, the average total EP rate is given in the leading order as 
\begin{equation}
 D^{-1}\langle \left(\mathbf{j}^{\rm irr}_\mathbf{p}\rho^{-1}\right)^2\rangle \simeq\gamma^{-1}\beta\langle\left(\gamma\mathbf{u}\right)^2\rangle=D_{\rm od}^{-1}\langle \left(\mathbf{j}_\mathbf{x}\rho(\mathbf{x},t)^{-1}\right)^2\rangle ~,
 \end{equation}
where $D_{\rm od}=(\beta\gamma)^{-1}$ is the diffusion coefficient for the overdamped dynamics. It is equal to the average total EP for the overdamped system.  

%%%%%%% Figure 1 %%%%%%%%%%%%%%%%%%%%%%%%%%%%%%%%
\begin{figure}
 \includegraphics*[width=\columnwidth]{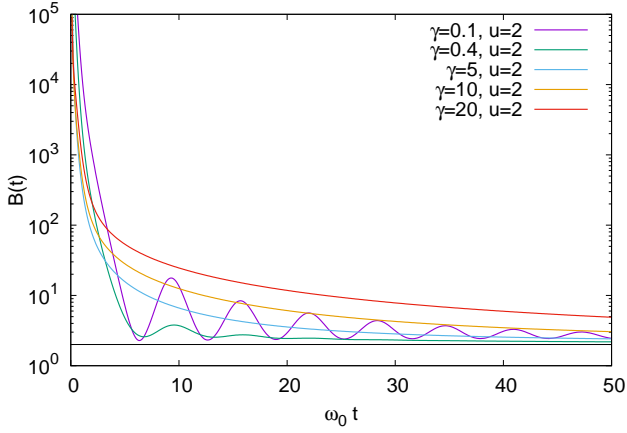}
\caption{$B(t)$ versus dimensionless time $\omega_0 t$ for the pulled harmonic oscillator with the angular frequency $\omega_0=\sqrt{k/m}$ above the horizontal line indicating the minimum bound $2$. Graphs are drawn from the analytic calculations for various values of $\gamma$ in unit of $\sqrt{km}$ and  pulling speed $u=2 $ in unit of $\omega_0(\beta k)^{-1/2}$.  For small $t$, $B(t)$ is divergent as $t^{-3}$ due to the scaling: ${\rm Var}W\sim t^2$, $EP\sim t$, and $DW\sim t^{3}$.}
\label{fig1}
\end{figure}
%%%%%% Figure 2 %%%%%%%%%%%%%%%%%%%%%%%%%%%%%%%%%
\begin{figure}
 \includegraphics*[width=\columnwidth]{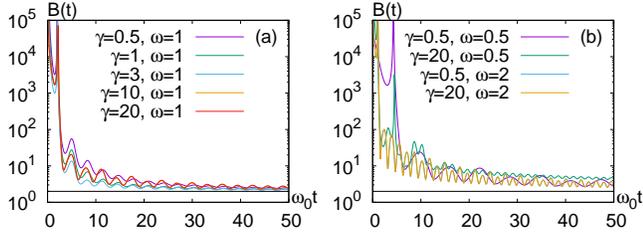}
\caption{$B(t)$ versus $\omega_0 t$ for the harmonic oscillator with the oscillating center above the horizontal line of the minimum bound $2$. Graphs are drawn from the analytical calculations for various values of $\gamma$ in unit of $\sqrt{km}$ and the frequency $\omega $ in unit of $\omega_0$. We use $A=1$, the root-mean-square (rms) distance $a=0.1$ in unit of $(\beta k)^{-1/2}$, and the rms velocity $v=0.1$ in unit of $(m\beta)^{-1/2}$, the equilibrium rms velocity. We use an initial inverse temperature $\beta'=0.5\beta$. The panel (a) is for $\omega=1$ and (b) for $\omega=0.5,~ 2$. For small $t$, $B(t)\sim t^{-3}$ because  ${\rm Var}W\sim t^2$, $EP\sim t$, and $DW\sim t^{3}$, which shows the same scaling behavior as in Fig.~\ref{fig1}. There are divergent peaks at specific times near $t=2$ for $\omega=1$, near $t=1$ for $\omega=2$, and near $t=4$ for $\omega=0.5$ with an additional peak near $t=7$ for large $\gamma=20$. It is due to the property that $DW=0$ at those times as it oscillates and grows from zero, which is characteristic of oscillating protocols and is not the case in Fig.~\ref{fig1}. }
\label{fig2}
\end{figure}
%%%%%%%Figure 3%%%%%%%%%%%%%%%%%%%%%%%%%%%%%%%
\begin{figure}
 \includegraphics*[width=\columnwidth]{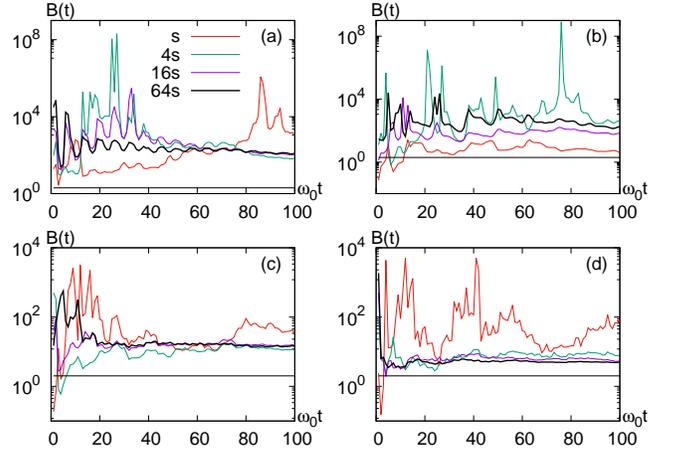}
\caption{ $B(t)$ versus $\omega_0t$ for the harmonic potential with the time-varying stiffness. Graphs are drawn from the computer simulations for (a) $\gamma=1,~\omega=1$ (b) $\gamma=2,~\omega=0.5$ (c) $\gamma=2,~\omega=1$ (d) $\gamma=2,~\omega=2$ in the same units as in Fig.~\ref{fig2}. We use $b=1/2$.  Graphs in each panel tend to approach to the black curve above the horizontal line of the minimum bound $2$ as the number of samples denoted by a multiple of $s=6.144\times 10^7$ increases. Due to large non-equilibrium fluctuations, relatively the huge number of samples are needed. Divergence and infinite peaks at small times shown in the previous figures are expected, but not clearly observed for the present numbers of samples used in the simulations. }
\label{fig3}
\end{figure}
%%%%%%%%%%%%%%%%%%%%%%%%%%%%%%%%%%%%%%%%%%

\emph{\textbf{Examples}}.
As the first example, we consider a particle in a harmonic trap potential of which the center is perturbed in time by an external device. The potential is given by $V(x, t)=k(x-a(t))^2/2$ for stiffness $k$. We consider the two cases: (i) $a(t)=u t$ for a constant pulling velocity $u$ playing the role of $\omega$ (ii) $a(t)=A \sin \omega t$. We investigate the TUR for work production $W$ given by $\int_0^\tau dt~ \partial_t V$ and the rate function $\Gamma(x,\omega t)=-ku(x-ut)$ for (i) and $-kA\omega\cos\omega t(x-A\sin\omega t)$ for (ii). An arbitrary initial state is given by the PDF $\propto e^{-(\beta'/2)(k(x-a)^2+(p-mv)^2/m)}$ for $\beta'\neq \beta$ where $a$ and $v$ are the average values of position and velocity at $t=0$. The problem is simplified as we separate variables such that $(x,p)=(z_x,z_p)+(X,P)$ for $X=\langle x\rangle$ and $P=\langle p\rangle$. Then, $\mathbf{z}=(z_x, z_p)$ is evolved by the equilibration process in the harmonic potential. Using $P=m\dot{X}$, $X$ satisfies $m\ddot{X}+\gamma \dot{X}+\alpha kX=ka(t)$ so $\bar{\mathbf{q}}=\mathbf(X,P)$ can be found easily. Then, the PDF for $\mathbf{q}=(x,p)$ is written as 
\begin{equation}
P(\mathbf{q},t)=\frac{1}{Z(t)}\exp\left[-\frac{1}{2}\mathbf{[\mathbf{q}-\overline{\mathbf{q}}(t)}]^{\rm t}\cdot\mathsf{A}(t)\cdot[\mathbf{\mathbf{q}-\overline{\mathbf{q}}(t)}]\right],
\end{equation}
where $Z(t)= \sqrt{(2\pi)^2/\det\mathsf{A}(t)}$. From our earlier study \cite{kwon_2011}, the covariance matrix of the Gaussian distribution is given by
$\mathsf{A}(t)^{-1}=\mathsf{A}_{\rm eq}^{-1}-e^{-\mathsf{F}t}\left[\mathsf{A}_{\rm eq}^{-1}-\mathsf{A}(0)^{-1}\right]e^{-\mathsf{F}^{\rm t} t}$~,
where $\mathsf{A}_{\rm eq}=\beta\left(\begin{array}{cc} k&0\\0&1/m\end{array}\right)$ and $\mathsf{F}=\left(\begin{array}{cc} 0&1/m\\k&\gamma/m\end{array}\right)$.
The case (i) has been extensively studied for the overdamped dynamics~\cite{vanzon1,vanzon2, kwangmoo} and the underdamped dynamics~\cite{kwon_jkps2018,kwon_3heats}. (ii) has recently been studied by our group  and  the detailed calculation is to be presented elsewhere \cite{youngchae}. We check the inequality in Eq.~\eqref{TUR_UND} for the work production $W$ by plotting 
\begin{equation}B(t)=\frac{({\rm Var} W)~EP}{DW^2}
\end{equation} 
where $EP=k_B^{-1}\langle\Delta S_{\rm tot}\rangle$ and $DW=-\gamma\partial_\gamma\langle W\rangle$. Figures~\ref{fig1} and \ref{fig2} shows that all $B(t)$'s for various values of parameters for $\gamma$ and $\omega$ ($u$) are above the horizontal line of $B(t)=2$. We confirm that Eq.~\eqref{D_Phi_new} holds for an arbitrary initial velocity $\mathbf{v}=v$ in one dimension.

We consider another type of the time-dependent harmonic potential given by a time-varying stiffness \cite{kwon_breathing,jun1,jun2}, given by $V=k(t)x^2/2$ with $k(t)=k(1+b\sin\omega t)$. The work production rate is equal to $\Gamma(x,\omega t)=\partial_t V(x, \omega t)=(kb\omega\cos\omega t) x^2/2$. This problem is investigated by using the computer simulation. Figure~\ref{fig3} shows that $B(t) \ge 2$ for various values of $\gamma$ and $\omega$.

\emph{\textbf{Summary}}. We derive the TUR for the product of the variance and EP for the time-accumulated production of an arbitrary observable in underdamped dynamics subject to general time-dependent protocols by exploiting the Cramers-Rao inequality. We show it to approach to the known TUR in overdamped dynamics for large viscosity limit. Any scaling function $s(\theta)$ with $s(0)=0$ and non-zero $s'(0)$ for the place of $(1+\theta)^{-1}$ in Eq.~\eqref{H_theta} goes to the unique TUR in the limit $\theta\to 0$, since ${\cal I}(\theta)$ and $(\partial_\theta\langle\Phi\rangle_\theta)^2 $ will have the same factor $s'(0)^2$. Divergence for small times and infinite-peak for oscillatory protocols present in the figures seem to be characteristic of the present TUR for short times, which might imply that the inequality bound is too loose. It is interesting to see whether there is an alternative TUR to give a tighter bound for small times.

We thank Professor Hyunggyu Park and Dr. Jongmin Park at Korean Institute of Advanced Studies for stimulating and helpful discussions. This work was supported by Basic Science Research Program of the National Research Foundation(NRF) funded by the Ministry of Education with Grant No. 2020R1A2C100976111 (CK) and 2018R1D1A1B07049254 (HKL).

\end{document}